\documentclass{article}

\usepackage{microtype}
\usepackage{graphicx}
\usepackage{subfigure}

\usepackage{hyperref}

\usepackage[accepted]{icml2019}

\icmltitlerunning{A Case for Backward Compatibility for Human-AI Teams}
\usepackage{booktabs} 
\usepackage{float}
\usepackage{todonotes}
\usepackage{enumitem}
\usepackage{bbm}
\usepackage{url}

\usepackage{soul}
\usepackage{relsize}

\usepackage[english]{babel}
\usepackage[markup=underlined]{changes}

\usepackage{todonotes}
\makeatletter
\setremarkmarkup{\todo[color=Changes@Color#1!20,size=\scriptsize]{#1: #2}}
\makeatother

\newcommand{\eg}{\mbox{\it e.g.}}
\newcommand{\ie}{\mbox{\it i.e.}}

\newcommand{\bi}{\begin{list}{$\bullet$}{
    \setlength{\leftmargin}{1.5 em}
    \setlength{\itemsep}{0 pt}
    \setlength{\topsep}{3 pt}
    \setlength{\parsep}{3 pt}
    \setlength{\partopsep}{0 pt}
    \setlength{\labelwidth}{1 em}
    \setlength{\labelsep}{0.5 em}
    \setlength{\parskip}{0cm}  }}
\newcommand{\ei}{\end{list}}

\newcommand{\BE}{\begin{enumerate}}
\newcommand{\EE}{\end{enumerate}}

\newcommand{\initab}{                           
\begin{tabbing}
XXX \= XXXX \= \kill
}
\newcommand{\begpub}{
\begin{quotation}
\noindent
}

\newcommand{\finpub}{
\end{quotation}
}

\newcommand{\name}{AI-advised human decision making}
\newcommand{\plat}{{\sc Caja}}
\newcommand{\hone}{\mbox{$h_1$}}
\newcommand{\htwo}{\mbox{$h_2$}}

\newcommand{\compatscore}{\mathcal{C}}
\newcommand{\loss}{L}
\newcommand{\lossbc}{\loss_c}
\newcommand{\lambdabc}{\lambda_c}
\newcommand{\dissonance}{\mathcal{D}}

\usepackage{amsmath}
\usepackage{amsthm}
\usepackage{amssymb}
\newtheorem*{definition}{Definition}
\newtheorem*{example}{Example}

\setitemize{noitemsep,topsep=0pt,parsep=0pt,partopsep=0pt, leftmargin=10pt}

\begin{document}

\twocolumn[
\icmltitle{A Case for Backward Compatibility for Human-AI Teams}



\icmlsetsymbol{equal}{*}
\begin{icmlauthorlist}
\icmlauthor{Gagan Bansal}{uw}
\icmlauthor{Besmira Nushi}{msr}
\icmlauthor{Ece Kamar}{msr}
\icmlauthor{Dan Weld}{uw}
\icmlauthor{Walter Lasecki}{um}
\icmlauthor{Eric Horvitz}{msr}

\icmlaffiliation{msr}{Microsoft Research}
\icmlaffiliation{uw}{University of Washington}
\icmlaffiliation{um}{University of Michigan}
\end{icmlauthorlist}

\icmlcorrespondingauthor{Gagan Bansal}{bansalg@cs.washington.edu}

\icmlkeywords{Machine Learning, Backward Compatibility, Human-AI Collaboration}

\vskip 0.3in
]
\printAffiliationsAndNotice{}



\begin{abstract}
AI systems are being deployed to support human decision making in high-stakes domains. In many cases, the human and AI form a team, in which the human makes decisions after reviewing the AI's inferences. A successful partnership requires that the human develops insights into the performance of the AI system, including its failures. We study the influence of {\em updates} to an AI system in this setting. While updates can increase the AI's predictive performance, they may also lead to changes that are at odds with the user's prior experiences and confidence in the AI's inferences, hurting therefore the overall \emph{team performance}. 
We introduce the notion of the {\em compatibility} of an AI update with prior user experience and present methods for studying the role of compatibility in human-AI teams. 
Empirical results on three high-stakes domains show that current machine learning algorithms do not produce compatible updates. We propose a re-training objective to improve the compatibility of an update by penalizing new errors. The objective offers full leverage of the performance/compatibility tradeoff, enabling more compatible yet accurate updates.
\end{abstract}

\section{Introduction}
A promising opportunity in AI is developing systems that can partner with people to accomplish tasks in ways that exceed the capabilities of either individually~\cite{wang2016deep,kamar2016directions,gaur2016effects}.
We see many motivating examples: a doctor using a medical expert system~\cite{wang2016deep}, a judge advised by a recidivism predictor, or a driver supervising a semi-autonomous vehicle. Despite rising interest, there is much to learn about creating effective human-AI teams and what capabilities AI systems should employ to be competent partners. 

We study human-AI teams in decision-making settings where a user takes action recommendations from an AI partner for solving a complex task. The user considers the recommendation and, based on previous experience with the system, decides to accept the suggested action or take a different action. We call this type of interaction {\em AI-advised human decision making}. The motivation for AI-advised human decision making comes from the fact that humans and machines have complementary abilities~\cite{wang2016deep,kamar2012combining} and that AI assistance can speed up decision making when humans can correctly identify when the AI can be trusted~\cite{lasecki2012scribe,lasecki2012real}. 

It might be expected that improvements in the performance of AI systems lead to stronger team performance, but, as with human groups, individual ability is only one of many factors that affect team effectiveness~\cite{dechurch-jap10,grosz1996collaborative}. In fact, the success of the team hinges on the human correctly deciding when to follow the recommendation of the AI system and when to override. If the human mistakenly trusts the AI system in regions where it is likely to err, catastrophic failures may occur. 
Human-AI teams become especially susceptible to such failures because of discrepancies introduced by system updates that do not account for human expectations. The following example illustrates this situation.


\begin{example}[\sc{Patient readmission}]
A doctor uses an AI system that is 95\% accurate at predicting whether a patient will be readmitted following their discharge to make decisions about enlisting the patient in a supportive post-discharge program. The special program is costly but promises to reduce the likelihood of readmission. After a year of interacting with the AI, the doctor develops a clear mental model that suggests she can trust the AI-advised actions on elderly patients. In the meantime, the AI's developer deploys a new 98\% accurate classifier, which errs on elderly patients. 
While the AI has improved by 3\%, the doctor is unaware of the new errors and might take the wrong actions for some elderly patients.
\vspace*{-0.2pc}
\end{example}

This example is motivated by real-world applications for reducing patient readmissions and other costly outcomes in healthcare~\cite{bayati2014data,caruana2015intelligible}, and motivates the need for reducing the cost of disruption caused by updates that violate  
users' mental models. Similar challenges are observed in other human-AI collaboration scenarios such as during over-the-air updates in the Tesla autopilot~\cite{tesla:2018}, and are present in a variety of other settings when AI services consumed by third-party applications, are updated. Despite these problems, developers almost exclusively optimize for AI performance.  Retraining techniques largely ignore important details about human-AI teaming, and the mental model that humans develop from interacting with the system. The goal of this work is to make the human factor a first-class consideration of AI updates. 
In summary, we make the following contributions: 

\begin{itemize}
\item We define the notion of compatibility of an AI update with the user's mental model created from past experience.
We then propose a practical adjustment to current ML (re)training algorithms --- an additional differentiable term to the logarithmic loss --- that improves compatibility during updates, 
and allows developers to explore the performance/compatibility tradeoff.

\item We introduce an open-source experimental platform
for 
studying (i) how people model the error boundary of an AI teammate for an AI-advised decision-making task, and (ii) how they adapt to updates. 

\item Using the platform, we perform user studies showing that updating an AI to increase accuracy, at the expense of compatibility, may {\em degrade} team performance. 
 Moreover, experiments on three high-stakes classification tasks (recidivism prediction, in-hospital mortality prediction, and credit-risk assessment) demonstrate that: (i) current ML models are not inherently compatible, but (ii) flexible performance/compatibility tradeoffs can be effectively achieved via a reformulated training objective.


\end{itemize}

\section{AI-Advised Human Decision Making}

In our studies, we focus on a simple, but common, model of human-AI teamwork that abstracts many real-world settings,
in which ML models support a human decision-maker. In this setting, which we call {\em \name}, an AI 
system provides a {\em recommendation}, but the human makes the final {\em decision}.
The team solves a sequence of tasks, repeating the following cycle for each time, $t$.

\begin{itemize}[leftmargin=.25in]
\item[S1:] The environment provides an input, $x^t$. 
\item[S2:] The AI (possibly mistaken) suggests an action, $h(x^t)$. 
 \item[S3:] Based on this input, the human makes her decision, $u^t$.
\item[S4:] The environment returns a reward, $r^t$, which is a function of the user's action, the (hidden) best action, and other costs of the human's decision (e.g., time taken).
\end{itemize}

\noindent While interacting over multiple tasks, the team receives repeated feedback about performance, which lets the human learn when she can trust the AI's answers. The cumulative reward $R$ over $T$ cycles records the team's performance.

\subsection{Trust as a Human's Mental Model of the AI}


\noindent  Just as for other automated systems~\cite{donald1988psychology}, humans create a mental model of AI agents~\cite{kulesza2012tell}. In \name, valid mental models of the reliability of the AI output improve collaboration by helping the user to know when to trust the AI's recommendation. A perfect mental model of the AI system's reliability could be harnessed to achieve the highest team performance. A simple definition for such a model would be $m: x \rightarrow \{T, F\}$, indicating which inputs the human trusted the AI to solve correctly. 
In reality, mental models are not perfect \cite{norman2014some}: users develop them through limited interaction with the system, and people have cognitive limitations.
Despite this, users learn and evolve a model of an AI system's competence over the course of many interactions. In the full version of this paper~\cite{bansal2019updates}, we show that these models can greatly improve team performance. In this study, we focus on the problem of updating an AI system in a way that it remains compatible with users' expectations.
\section{Compatibility of Updates to Classifiers}

In software engineering, an update is \emph{backward compatible} if the updated system can support legacy software.
By analogy, we define that an update to an AI component is {\em locally compatible} with a user's mental model if it does not introduce new errors and the user, even after the update, can safely trust the AI's recommendations. 

\begin{definition}[\textsc{Locally-Compatible Update}]
Let $m(x)$ denote a mental model that dictates the user's trust of the AI on input $x$.  Let $A(x,u)$ denote whether $u$ is the appropriate action for input $x$. 
An update, \htwo, to a learned model, \hone, is locally compatible with $m$ iff 
\[\forall x, [m(x) \wedge A(x, \hone(x))] \Rightarrow A(x, \htwo(x)) \]
\end{definition}

In other words, an update is compatible only if, for every input where the user trusts the AI and \hone\ recommends the correct action, the updated model, \htwo, also recommends the correct action. In the rest of this paper, we focus on situations where a classifier's predictions are actions. For instance, in the patient readmission example, if a classifier predicts that the patient will be readmitted in the next 30 days, the suggested action from the classifier would be to include the patient in a special post-discharge program.

\subsection{Globally Compatible Updates}

When developers are building an AI system that is used by many individuals, it may be too difficult to track individual mental models or to deploy different updated models to different users. In this situation, an alternative to creating locally compatible updates, is a {\em globally compatible update}. 
\begin{definition}[\textsc{Globally-Compatible Update}]
An updated model, \htwo,  is globally compatible with \hone, iff
\[\forall x, A(x, \hone(x))\Rightarrow A(x, \htwo(x)) \]
\end{definition}

Note that a globally compatible update is locally compatible for {\em any} mental model.  While global compatibility is a nice ideal, satisfying it for all instances is difficult in practice. More realistically, we seek to minimize the number of errors made by \htwo that were not made by $\hone$, since that will hopefully minimize confusion among users. 
To make this precise, we introduce the notion of a {\em compatibility score}.

\begin{definition} [\textsc{Compatibility Score}]   
The compatibility score $\compatscore$ of an update $\htwo$ to $\hone$ is given by the fraction of examples on which $\hone$ recommends the correct action, $\htwo$ also recommends the correct action.
\begin{equation}
\label{eq:score}
    \compatscore(\hone, \htwo) =  \frac{\sum_{x}A(x, \hone(x)) \cdot A(x, \htwo(x))}{\sum_{x }A(x, \hone(x))}
\end{equation}
\end{definition}
If $\htwo$ introduces no new errors, $\compatscore(\hone, \htwo)$ will be 1. Conversely, if all the errors are new, the score will be 0. 

\subsection{Dissonance and Loss}
To train classifiers, ML developers optimize for the predictive performance of $\htwo$
by minimizing a classification loss $\loss$ that penalizes low performance. The equation below shows the negative logarithmic loss for binary classification -- a commonly used training objective in ML. 
\begin{equation*}
    \label{eq:logloss}
    \loss(x, y, \htwo) = y \cdot \log p(\htwo(x)) + (1 - y) \cdot \log (1 - p(\htwo(x)))
\end{equation*}
Here, the probability $p(h(x))$ denotes the confidence of the classifier that recommendation $h(x)$ is true, while $y$ is the true label for $x$ (\ie, $A(x, y) = \mbox{\em True}$).
The negative log loss, like many other loss functions in machine learning, depends only on the true label and the confidence in prediction -- it ignores the previous versions of the classifier and, hence, has no preference for compatibility. As a result, retraining using different data can lead to very different hypotheses, introduce new errors, and decrease the compatibility score. 
To alleviate this problem, we define a new loss function $\lossbc$ expressed as the sum of classification loss and {\em dissonance}.

\begin{definition}[\textsc{Dissonance}]
The dissonance $\dissonance$ of $\htwo$ to $\hone$ is a function $\dissonance: x, y, \hone, \htwo \rightarrow \mathcal{R}$ that penalizes a low compatibility score. Furthermore, $\dissonance$ is differentiable.
\begin{equation}
    \label{eq:diss}
    \dissonance(x, y, \hone, \htwo) = \mathbbm{1}(\hone(x) = y) \cdot \loss(x, y, \htwo)
\end{equation}
\end{definition}

\noindent Recall that  $\compatscore(\hone, \htwo)$ is high when both $\hone$ and $\htwo$ are correct (Eqn~\ref{eq:score}).  Dissonance  expresses the opposite notion: measuring if $\hone$ is correct ($\mathbbm{1}$ denotes an indicator function)
and penalizing by the degree to which $\htwo$ is incorrect.
Equation~\ref{eq:lossbc} defines the new loss.
\begin{equation}
    \label{eq:lossbc}
    \lossbc = \loss + \lambdabc \cdot \dissonance
\end{equation}

\noindent Here, $\lambdabc$ encodes the relative weight of dissonance, controlling the additional loss to be assigned to all new errors.
We refer to this version as {\em new-error dissonance}.
Just as with classification loss, there are other ways to realize dissonance. We explored two alternatives, which we refer to as {\em imitation} and {\em strict imitation} dissonance.
Eqn~\ref{eq:diss_imitation} describes the imitation dissonance which measures the log loss between the prediction probabilities of $\hone$ and $\htwo$:
\begin{equation}
    \label{eq:diss_imitation}
    \dissonance'(x, y, \hone, \htwo) = \loss(x, \hone, \htwo)
\end{equation}

\noindent Eqn~\ref{eq:diss_imitation} is used in model distillation~\cite{ba2014deep,hinton2015distilling}, where the aim is to train a shallower, less expensive model by imitating the probabilities of larger, accurate model. Unfortunately, $\dissonance'$ has the effect of nudging \htwo\ to mimic \hone's mistakes as well as its successes.  
Eqn~\ref{eq:diss_imitation_strict} describes the strict imitation dissonance, which follows a similar intuition but it only adds the log loss between $\hone$ and $\htwo$ when $\hone$ is correct.
\begin{equation}
    \label{eq:diss_imitation_strict}
    \dissonance''(x, y, \hone, \htwo) = \mathbbm{1}(\hone(x) = y) \cdot \loss(x, \hone, \htwo)
\end{equation}

\section{Platform for Studying Human-AI Teams}
\begin{figure}[t]
    \centering
    \includegraphics[width=0.8\linewidth]{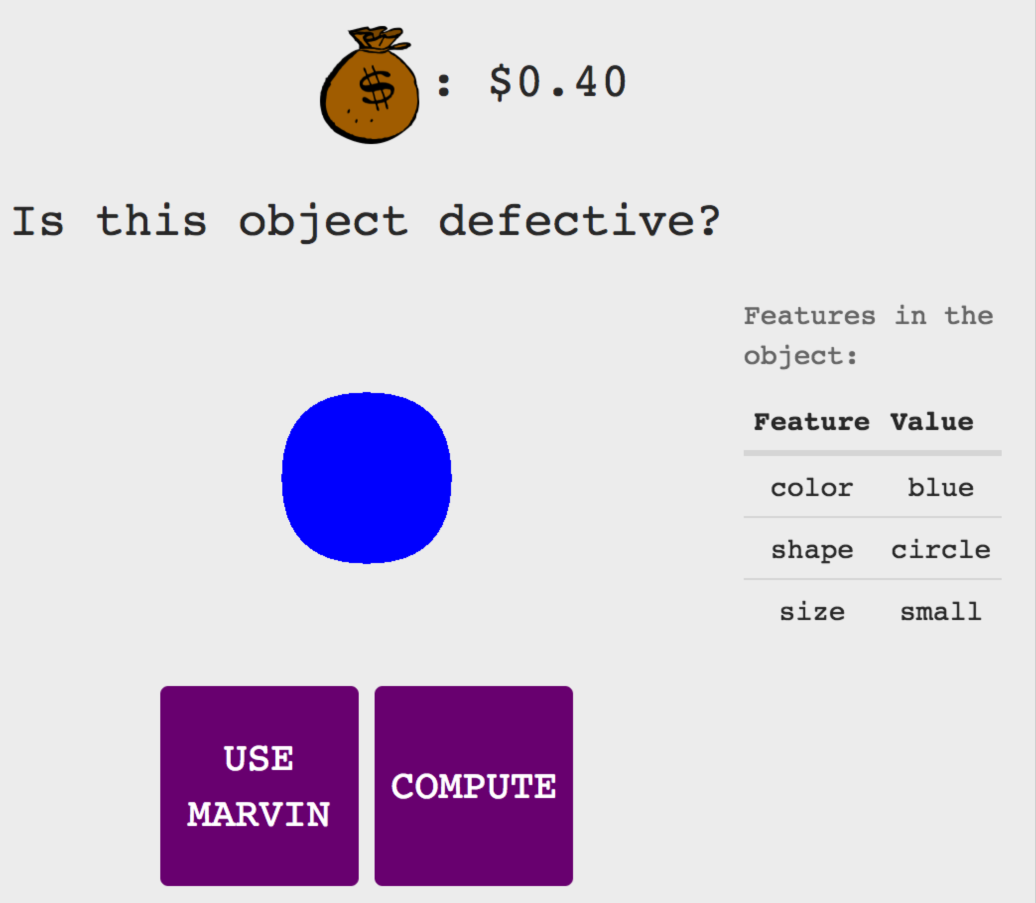}
    \caption{\plat\ platform for studying human-AI teams.}
    \label{fig:ui}
\end{figure}  

How might we study the impact of AI accuracy, updates, compatibility, and mental models on the performance of \name\ teams? Ideally, we would conduct user studies in real-world settings, varying parameters like the length of interaction, task and AI complexity, reward function, and the AI's behavior. However, testing in real settings reduces or removes our ability to directly control the performance of the AI and it may largely measure experts' differing experience in the domain, rather than their interactions with the AI. To control for human expertise and the centrality of mental modeling, we developed the \plat\ platform, which supports parameterized user studies in an assembly line domain that abstracts away the specifics of problem solving and focuses on understanding the effect of mental modeling on team success. \plat\ is designed such that {\em no} human is a task expert (nor can they become one). In fact, the true label of decision problems is randomly generated so that people cannot learn how to solve the task. However, humans can learn when their AI assistant, Marvin, succeeds and when it errs. Alongside, the human has access to a perfect problem-solving mechanism, which she can use (at extra cost) when she does not trust Marvin. 

\plat\ is a web-based game, whose goal is to make classification decisions for a fixed number of box-like objects.  For each object, the team follows the steps S1-S4 to decide whether the object is ``defective" or not. In S1 a new object appears (\eg,\ blue square), in S2 the AI recommends a label (\eg,\ not-defective), in S3 the player chooses an action (\eg,\ accept or reject the AI recommendation), and in S4 the UI returns a reward and increments the game score. The objects are composed of many features, but only a subset of them are made {\em human-visible}.
For example, visual properties like shape, color, and size are visible, but the contents are not. In contrast, the AI has access to all the features but 
may make errors. At the beginning of the game, users have no mental model of the AI's error boundary. However, to achieve high scores, they must learn a model using feedback from step S4. Figure~\ref{fig:ui} shows a screenshot of the game at step S3. 

\plat\ allows study designers to vary parameters, such as the number of objects, number human-visible features, reward function, AI accuracy, and complexity of perfect mental model (number of clauses and literals in the   error boundary and stochasticity of errors). Further, it enables one to study the effects of updates to AI by allowing changes to these parameters at any time step. In the next section, we use \plat\ to answer various research questions.

\section{Experiments}
We present experiments and results in two parts. First, using our platform, we conduct user studies to understand the impact of updates on team performance. Second, we simulate updates for three real-world, high-stakes domains and show how the retraining objective enables an explorable tradeoff between compatibility and performance.

\begin{table}[t]
    \centering
    \begin{tabular}{|c|c|c|}
    \hline
         & Accept & Compute  \\
         \hline
         AI right & \$0.04 & 0 \\
         \hline
         AI wrong & -\$0.16 & 0\\
         \hline
    \end{tabular}
    \caption{Reward matrix for the user studies. To mimic high-stakes domains, penalty for mistakes is set to high. 
    }
    \label{tab:payoff}
\end{table}

\subsection{User Studies} 
In user studies, we hired MTurk workers and directed them to the \plat\ platform. The task of workers was to form a team with an AI, named Marvin, and label a set of objects as ``defective'' or ``not defective''.
Following \name, to label an object, a worker can either accept Marvin's recommendation, which is initially correct 80\% of the time, or use the ``compute'' option, which is a surrogate for the human doing the task herself perfectly but incurring an opportunity cost.

Table~\ref{tab:payoff} summarizes the reward function  used in our studies. The matrix is designed in a way that it imitates a high-stakes scenario, i.e., the monetary penalty for a wrong decision is much higher than the reward for a correct decision. Note that the expected value of a naive strategy (\eg, always ``Compute"  or always ``Accept," without considering the likelihood of Marvin's correctness) is zero. The only way to get a higher score is by learning when to trust Marvin by playing the game. An error boundary, $f$, is expressed as a conjunction of literals. For example, one possible error boundary would be $f = (blue \cap square)$, which means that Marvin errs at all objects that are blue and have a squared shape. Since many features can be used as literals, we chose them randomly to create isomorphic error boundaries. In the full paper~\cite{bansal2019updates}, we show that it is easier for workers to create a mental model for boundaries that have fewer literals and are not \emph{stochastic}. Next, we show results from experiments that introduce updates.

\begin{figure}[t]
  \centering
    \includegraphics[width=\columnwidth]{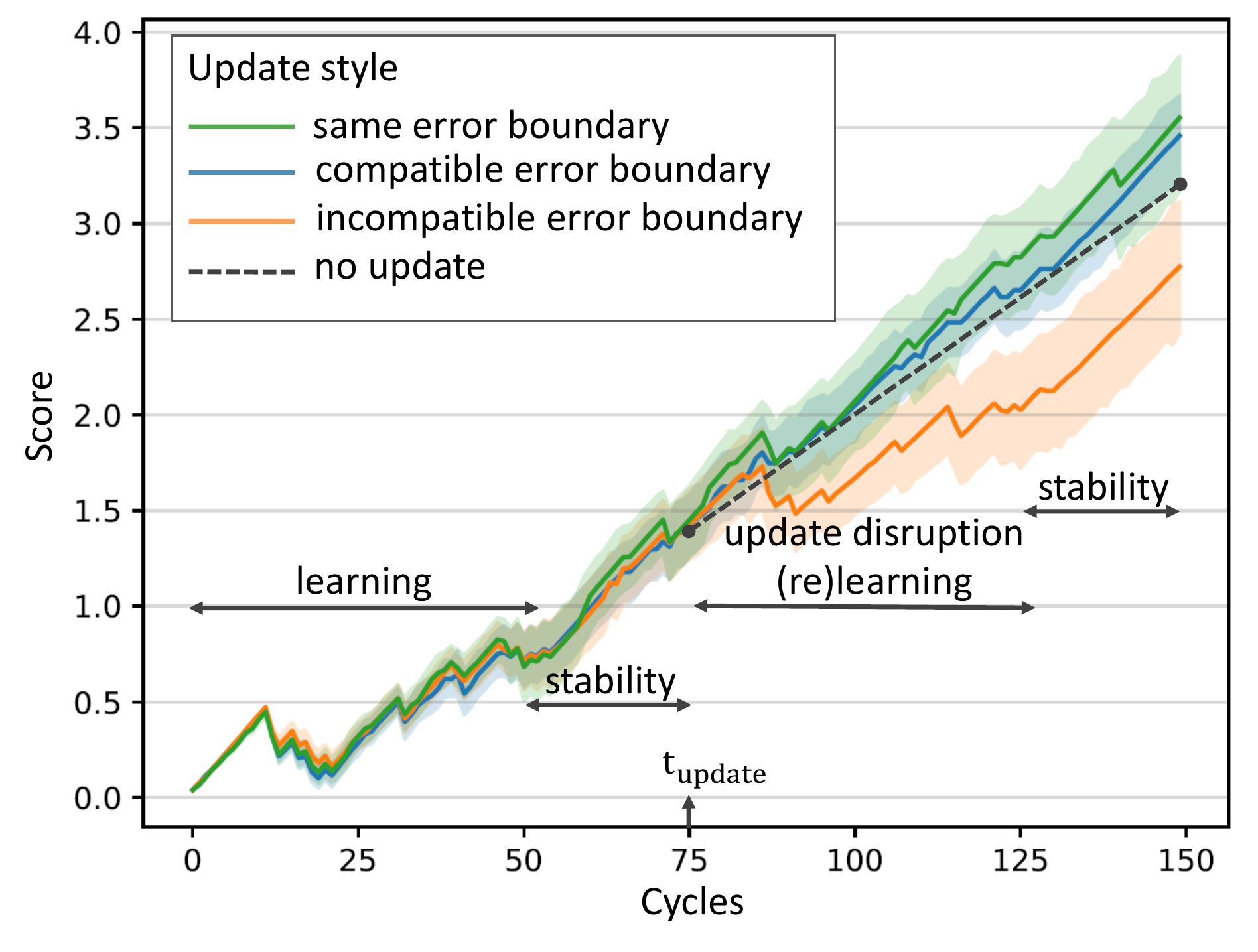}
    \caption{Team performance for different update settings. Compatible updates improve team performance, while incompatible updates hurt team performance  despite improvements in AI accuracy.}
    \label{fig:updateExp}
\end{figure}

\noindent {\bfseries Q1: }{\em Do more compatible updates lead to higher team performance than incompatible updates?}\\
\noindent To study the impact of updates, we set the number of cycles to 150, and at the 75th cycle, update the classifier to a version that is 5\% more accurate (80\% $\rightarrow$ 85\%). Then, we divide the participants into three groups: same error boundary,
compatible error boundary, and incompatible error boundary. The same error boundary group receives an update improving accuracy, but the  error boundary is unchanged. For the two other groups, the number of literals (features) in the error boundary changes from two to three. The update for the compatible error boundary group introduces no new errors; for example, if before the update the error boundary was $blue \cap square$, after the update it may change to $small \cap blue \cap square$. For the incompatible error boundary group, the error boundary introduces new errors violating compatibility. Figure~\ref{fig:updateExp} summarizes our results. We also show the performance of workers if no update was introduced (dashed line). The graph demonstrates two main findings on the importance of compatibility. First, a more accurate but incompatible classifier  results in lower team performance than a less accurate but compatible classifier (no update) because workers have to relearn the new incompatible error boundary. Second, compatible updates improve team performance. Moreover, the figure shows different stages during the interaction: the user learning the original error boundary, team stabilizes, update causes disruption, and performance stabilizes again. 

\subsection{Experiments with  High-Stakes Domains}
\noindent {\bfseries Datasets.} To investigate whether a tradeoff exists between  performance and compatibility of an update, we simulate updates to classifiers for three domains: recidivism prediction (Will a convict commit another crime?)~\cite{angwin2016machine}, in-hospital mortality prediction (Will a patient die in the hospital?)~\cite{johnson2016mimic,harutyunyan2017multitask}, and credit risk assessment (Will a borrower fail to pay back?)\footnote{\url{https://community.fico.com/s/explainable-machine-learning-challenge}}. We selected these high-stakes domains to highlight the potential cost of mistakes caused by incompatible updates in human-AI teams.

\begin{table}[t]
\footnotesize
\centering
\begin{tabular}{|l|l|c|c|c|}
\hline
Classifier & Dataset                         & ROC $\hone$ & ROC $\htwo$ & $\compatscore(\hone, \htwo)$ \\
\hline
LR & Recidivism       &   0.68 &        0.72     &   0.72               \\
& Credit Risk          &   0.72     &    0.77    &   0.66              \\
& Mortality    &    0.68 &    0.77    &    {\bf 0.40}                    \\
\hline
MLP & Recidivism   &   0.59     &     0.73   &     0.53                    \\
& Credit Risk     &  0.70   &    0.80    &     0.63                     \\
& Mortality &       0.71 &     0.84   &     0.76            \\
\hline
\end{tabular}
\caption{\label{tab:score} Although training on more data increases classifier performance, compatability can be suprisingly low.}
\end{table}

\begin{figure*}[t]
    \centering
    \includegraphics[width=0.85\linewidth]{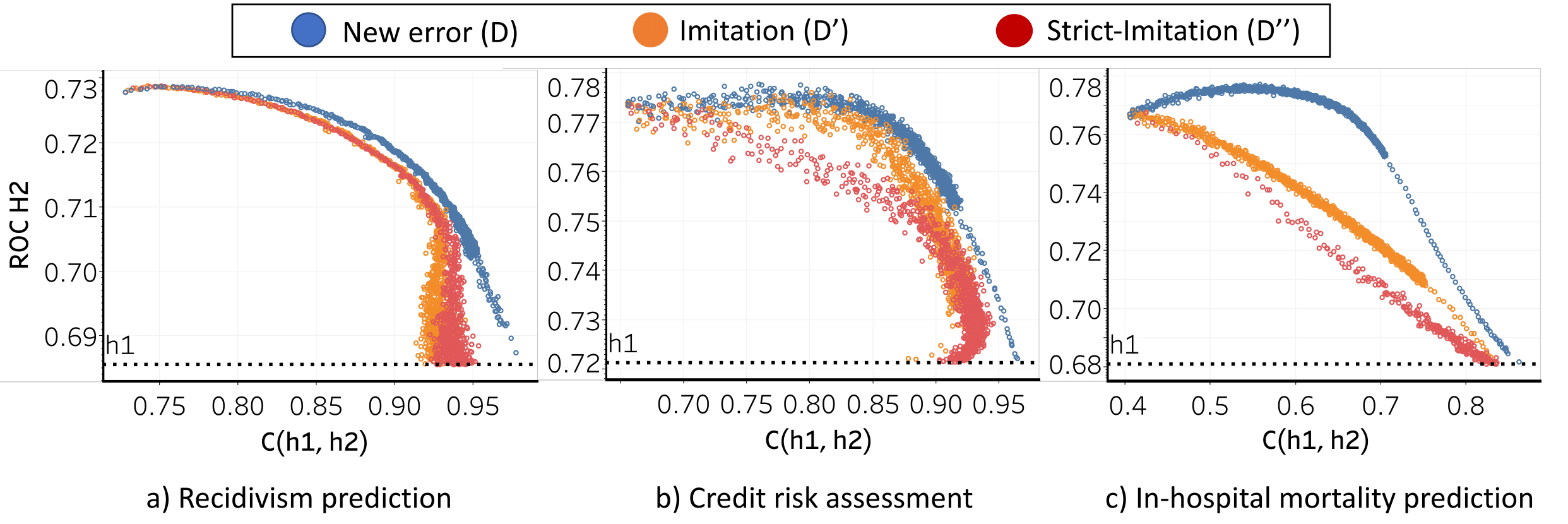}
    \vspace{-0.6em}
    \caption{Performance vs. compatibility for a logistic regression classifier. The reformulated training objective ($\lossbc$) offers an explorable performance/compatibility tradeoff, generally more forgiving during the first half of the curves. The training objective based on new-error dissonance performs the best, whereas the ones based on imitiation and strict-imitation dissonance perform worse since they imitate probabilities of a less accurate, and less calibrated model ($\hone$).}
    \label{fig:diss}
\end{figure*}
\noindent {\bfseries Q2: }{\em Do ML classifiers produce compatible updates?}\\
\noindent For this experiment, we first train a classifier $\hone$ on 200 examples and note its performance. Next, we train another classifier $\htwo$ on 5000 examples
and note its performance and compatibility score. We train both classifiers by minimizing the negative log loss. Table~\ref{tab:score} shows the performance (area under ROC) and compatibility averaged over 500 runs for logistic regression (LR) and multi-layer perceptron (MLP) classifiers. We find that training $\htwo$ by just minimizing log loss does not ensure compatibility. For example, for logistic regression and the in-hospital mortality prediction task, the compatibility score is as low as 40\%.  That is, 60\% of the instances where $h_1$ was correct are now violated.

\noindent {\bfseries Q3: }{\em Is there a tradeoff between the performance and the compatibility of an update to AI?}\\
\noindent For Q3 (and Q4), we learn the second classifier $\htwo$ by minimizing $\lossbc$. As $\lossbc$ depends also on the first classifier, we make its prediction available to the learner. We vary $\lambdabc$ and summarize the resulting performance and compatibility scores across different datasets for the logistic regression classifier in Figure~\ref{fig:diss} and for different definitions of dissonance (the full paper also provides results for MLP). 
The figure shows that there exists a tradeoff between the performance of $\htwo$ and its compatibility to $\hone$. This tradeoff is generally more flexible (flat) in the first half of the curves. This shows that, at the very least, one can choose to deploy a more compatible update without significant loss in accuracy. Although such updates are not fully compatible, they might still be relevant to be picked by the developer if the update is supported by efficient explanation techniques that can help users to better understand how the model has changed. In these cases, a more compatible update would also reduce the effort of user (re)training. In the second half, the tradeoff becomes more evident. High compatibility can sacrifice predictive performance. Look-up summaries similar to graphs shown in Figure~\ref{fig:diss} are an insightful tool for ML developers that can guide them select an accurate yet compatible model based on the specific domain requirements. 

\noindent {\bfseries Q4: }{\em What is the relative performance of the different dissonance functions?}\\
 Figure~\ref{fig:diss} compares the performance of the new-error dissonance function ($\dissonance$) with the imitation-based dissonances ($\dissonance'$  and  $\dissonance''$). As anticipated, $\dissonance$ performs  best on all  three domains. The definitions inspired by model distillation, $\dissonance'$ and $\dissonance''$,  assume that \hone\ is calibrated, and more accurate. Therefore, $\htwo$ needs to remain faithful to only the correct regions of a less accurate model $\hone$. If these assumptions are violated, $\htwo$ overfits to non-calibrated confidence scores of $\hone$, which hurts performance.
\section{Discussion and Directions}


This work showed that backward compatibility is an essential determinant of team performance, and developers should factor it in system design supported by guiding tools exploring the performance/compatibility tradeoff in a similar way as shown in Figure~\ref{fig:diss}. Varying $\lambdabc$ results in numerous models on the performance/compatibility spectrum. The decision to select the appropriate model depends on several factors, including the user ability to create a mental model, the cost of disruption, and the availability of other alternative approaches for minimizing disruption caused by updates. One such approach is to retrain the user, for example, by leveraging mechanisms from interpretable AI to explain the updated model to users or to explain differences between \hone\ and \htwo. However, this may not always be practical: 
(1) in practice, developers may push updates frequently, and since re-training requires user’s additional time and effort, it may not be practical to subject experts to repeated re-training; 
(2) updates can arbitrarily change the decision boundary of a classifier, and as a result, require the user to re-learn a large number of changes;
(3) re-training requires the developers to create an effective curriculum or generate a “change summary” based on the update. It is often impossible to compute such summaries in a human-interpretable way.
Nevertheless, backward compatibility does not preclude retraining; these techniques are complementary to each other. 

Another complementary approach is to share the AI's confidence in the prediction. Well-calibrated confidence scores can help a user to decide when or how much to trust the system. Unfortunately, confidence scores of ML classifiers are often not calibrated~\cite{nguyen2015deep} or a meaningful confidence definition may not exist.


\section{Related Work}
Prior fundamental work explored the importance of mental models for achieving high performance in group work~\cite{grosz1999evolution}, human-system collaboration \cite{rouse1992role}, and interface design~\cite{carroll1988mental}. Our work builds upon these foundations and studies the problem for \name.
Previous work \cite{zhou2017effects} investigated factors that affect user-system trust, e.g., model uncertainty and cognitive load. The platform proposed in this work enables human studies that can analyze the effect of such factors.

The field of software engineering has initially studied the problem of backward compatibility, seeking to design components that remain compatible with a larger software ecosystem after updates~\cite{bosch2009software,spring2005techniques}.
Machine learning research has explored related notions. {\em Stability} expresses the ability of a model to not significantly change its predictions given  small changes in the training set~\cite{bousquet2001algorithmic}. {\em Consistency}, which has application in ML fairness, is a property of smooth classifiers, which output similar predictions for similar instances~\cite{zhou2004learning}. 
\emph{Catastrophic forgetting} is an anomalous behavior of neural network models that occurs when they are sequentially trained to perform multiple tasks and forget to solve earlier tasks over time~\cite{kirkpatrick2017overcoming}. 
While  these concepts are fundamental for analyzing changing trends in continuously learned models, they do not consider human-AI {\em team} performance nor prior user experience. Related to our proposed retraining objective is the idea of \emph{cost-sensitive} learning~\cite{elkan2001foundations}, where different mistakes may cost differently; for example, false positives may be especially costly. In our case, the cost also depends on the behavior of the previous model $\hone$.

\section{Conclusions}
    We studied how updates to an AI system can affect human-AI team performance and introduced methods and measures for characterizing and addressing the compatability of updates. We introduced \plat, a platform for measuring the effect of AI performance and the effect of updates on team performance. Since humans have no experience with \plat's abstract game, the platform controls for human problem-solving skill, distilling the essence of mental models and trust in one's AI teammate.
    Using \plat, we presented experiments  demonstrating how an update that makes an AI component more accurate can still lead to diminished human-AI team performance. 
    We introduced a practical re-training objective that can improve the compatibility of updates. 
    Experiments across three data sets show that our  approach creates updates that are more compatible, while maintaining high accuracy. 
    Therefore, at the very least, a developer can choose to deploy a more compatible model without sacrificing performance. 

\clearpage
\bibliography{main}
\bibliographystyle{icml2019}
\end{document}